# Urban Exodus? Understanding Human Mobility in Britain During the COVID-19 Pandemic Using Facebook Data

Francisco Rowe[1,*], Alessia Calafiore[1], Daniel Arribas-Bel[1,2], Krasen Samardzhiev[1], MartinFleischmann[1]

[1] Department of Geography and Planning, University of Liverpool, Liverpool, United Kingdom, L69 7ZT.
[2] The Alan Turing Institute, British Library, 96 Euston Road, London, England, NW1 2DB, United Kingdom

*Corresponding author: F.Rowe-Gonzalez@liverpool.ac.uk

**Abstract**
Existing empirical work has focused on assessing the effectiveness of non-pharmaceutical interventions on human mobility to contain the spread of COVID-19. Less is known about the ways in which the COVID-19 pandemic has reshaped the spatial patterns of population movement within countries. Anecdotal evidence of an urban exodus from large cities to rural areas emerged during early phases of the pan- demic across western societies. Yet, these claims have not been empirically assessed. Traditional data sources, such as censuses offer coarse temporal frequency to analyse population movement over short-time intervals. Drawing on a data set of 21 million observations from Facebook users, we aim to analyse the extent and evolution of changes in the spatial patterns of population movement across the rural-urban continuum in Britain over an 18-month period from March, 2020 to August, 2021. Our findings show an overall and sustained decline in population movement during periods of high stringency measures, with the most densely populated areas reporting the largest reductions. During these periods, we also find evidence of higher-than-average mobility from highly dense population areas to low densely populated areas, lending some support to claims of large-scale population movements from large cities. Yet, we show that these trends were temporary. Overall mobility levels trended back to pre-coronavirus levels after the easing of non-pharmaceutical interventions. Following these interventions, we also found a reduction in movement to low density areas and a rise in mobility to high density agglomerations. Overall, these findings reveal that while COVID-19 generated shock waves leading to temporary changes in the patterns of population movement in Britain, the resulting vibrations have not significantly reshaped the prevalent structures in the national pattern of population movement.

**Keywords**: Human mobility; Internal migration; Mobile phone location data; Facebook; COVID-19; Urban exodus; Great Britain

# 1. INTRODUCTION

The COVID-19 pandemic has led to major changes in the patterns of human mobility within and between countries. In addition to stringent international travel restrictions, non-pharmaceutical interventions to contain the spread of COVID-19, such as lock- downs, social distancing measures, school and business closures, have transformed daily life behaviours resulting in reduced overall levels of mobility (Nouvellet et al. 2021). Especially during lockdowns, mobility recorded reductions in the frequency, distance and time of trips across the world (e.g. Department for Transport 2021; Mobility 2021). Rises in teleworking, online schooling and remote shopping activity reduced the need to travel for work, education, shopping and leisure. Coupled to fear of crowded public spaces, non-pharmaceutical interventions also prompted more geographically localised mobility patterns (Engle, Stromme, & Zhou 2020; Linka, Goriely, & Kuhl 2021), and a modal shift away from mass public transit to private, active (notably walking) and e-forms of mobility, such as e-biking and e-scooting (Li et al. 2021; Mobility 2021).

These changes and fears have generated a passionate debate about the future of big cities. Some have predicted that COVID- 19 would create a tipping point leading to "the end of cities", while others have made predictions anticipating strong urban recovery and resilience. Building on current knowledge and past pandemics, studies have carefully analysed existing evidence and anticipated the potential immediate and long-term economic and social reverberations of the COVID-19 pandemic on the structure and morphology of cities and regions (Florida, Rodríguez-Pose, & Storper 2021; Nathan 2020; Sharifi & Khavarian- Garmsir 2020). These changes are anticipated to take place at a microgeographical scale, altering the organisation of people and activity within urban regions, between central and suburban areas, and at fine granular spaces, such as neighbourhoods and districts (Florida et al. 2021). Yet, these changes are not expected to significantly reshape existing structural national patterns of population settlement and economic systems at a macrogeographical scale (Florida et al. 2021).

During the early phases of the pandemic, reports of an "urban exodus" emerged with rampant speculation that this trend would persist post-COVID-19. Big cities became the epicentres of COVID-19 infection during the first wave of the pandemic. Coupled with slow government responses to contain community transmission, high population density, air connectivity and spatial concentration of public-facing jobs contributed to the clustering of COVID-19 cases in large cities during early stages of the pandemic in many western countries (Florida et al. 2021; Nathan 2020). These factors facilitate the spread of COVID- 19 through the creation of dense networks of social interaction (Chang et al. 2021). As we were learning about the severity of COVID-19 infection, cases spread throughout the world, deaths rose and teleworking went mainstream during the early parts of 2020, anecdotal evidence suggests that large numbers of city dwellers sought to leave large cities to avoid crowded places and in the pursuit of more personal space and access to natural amenities (Nathan & Overman 2020; Specia 2021; Whitaker 2021).

Anecdotal evidence has also indicated that big city residents moved to second homes, vacation towns, suburbs and nearby, smaller, less expensive locations (Paybarah, Bloch, & Reinhard 2020; Quealy 2020). Adding to the existing pressures from climate change, air pollution, high housing costs and urban crime, the pandemic has brought about changes allowing non-work related considerations to take a more important role in people's residential choices (Nathan 2020). In large cities, lockdown restrictions have confined households to costly small living spaces, with limited capacity to accommodate multi home-office-school-leisure functionalities. Business closures and social distancing measures stripped away the effervescence of social interaction, busy and dense urban spaces, and vibrant leisure activities (Florida et al. 2021). Additionally, remote work has reduced the need for frequent commuting, and hence, living close to workplaces. Coupled to access to green and open spaces, larger residential living and less crowded spaces thus seem to have become more prominent residential preferences since the start of the pandemic.

While these changes are expected to alter micro-level household decision choices, they are not expected to significantly reshape the existing macro-level patterns of national population settlement and economic systems (Reades & Crookston 2021). Big cities have successfully weathered previous pandemics (Acuto 2020; Glaeser 2020) and are likely to remain attractive places to live. Agglomeration economies are an essential feature of cities. Cities facilitate the clustering of talent and economic assets, consumer base, face-to-face interaction and diversity that are key to fostering innovation, creativity and economic growth (Storper & Venables 2004). New forms of hybrid work are likely a permanent outcome of the pandemic (Kniffin et al. 2021). Telework is a poor substitute for high-contact, knowledge-based work. At the same time, most remote locations do not have the digital infrastructure and diversity of services required to cater for city residents - and initiatives are already underway to make cities more resilient to future pandemics (Bereitschaft & Scheller 2020; Florida & Pedigo 2020; Kraus & Koch 2021).

Thus, while we have learned about changes in human mobility within cities in relation to non-pharmaceutical interventions, less is known about the patterns of population redistribution across the national territory during the COVID-19 pandemic. A key challenge has been the lack of suitable data to capture national-scale patterns of population movements across the rural- urban continuum as the pandemic evolves. Traditionally census and population register data have been used to explore human mobility patterns at such scale. However, these data systems are not regularly updated, lacking the temporal granularity to analyse population movements over short-time intervals. Digital traces data derived from mobile phone applications now provide a unique opportunity to capture these movements at an unprecedented spatial and temporal granularity (Green, Pollock, & Rowe 2021).

Drawing on Facebook users' mobile phone location data, this paper aims to analyse the extent and durability of changes in human mobility patterns across the rural-urban continuum in Britain during the COVID-19 pandemic, covering an 18-month period from March 2020 to August 2021. Specifically this paper seeks to address the following set of questions:

- To what extent have people moved away from cities, and redistributed across the urban-rural continuum during the pandemic?
- What have been the key interactions between places across the population density hierarchy? Have people mainly moved to sparely populated areas?
- To what extent the intensity of population movement from cities have been sustained throughout the pandemic? Have the observed changes been temporary, or are likely to persist post pandemic?

The rest of the paper is structured as follows. The next section reviews the emerging evidence and hypotheses about the spatial patterns of population movements from cities during the COVID-19 pandemic, before discussing the predominant trends of population movements in Britain in the years preceding the pandemic. Section 3 describes the data, and Section 4 discusses the methods used in this study. Section 5 presents the key results from our analyses before they are discussed in the light of the existing literature in Section 6, which also identifies key limitations and potential avenues for future work.

## 2. BACKGROUND

*2.1 Emerging evidence on mobility patterns across the urban hierarchy during COVID-19*

As COVID-19 expanded throughout the world in February and March in 2020, anecdotal evidence of an "urban exodus" from big cities emerged in many western societies (Sagnard 2021; Weeden 2020). At early stages of the pandemic, little was known about the virus, and globally connected cities were hit hardest (Florida et al. 2021; Matheson et al. 2020). By November 2020, approximately 95% of all the reported infections and fatalities had occurred in a few large cities

(Pomeroy & Chainey 2020). Newspapers' headlines were speculating about "the end of cities" (e.g. Pomeroy & Chainey 2020). In the UK, reports based on data from the property website Rightmove indicated that the number of online inquiries from residents living in the ten largest cities looking for a village property was reported to increase by 126% in June-July 2020, relative to the same period in 2019 (Marsh 2020). In France, increases in real estate transactions outside cities were also linked to city residents moving to smaller towns or villages (Sagnard 2021). In the US, a rise of 30 percentage points in the number of households moving from large metropolitan areas was reported based on data from mail-forwarding requests and credit report data (Paybarah et al. 2020). In Australia, the Australian Bureau of Statistics estimated a net loss of 11,000 people from capital cities during the September quarter of 2020 (Davies 2021).

COVID-19 has exposed key imperfections of living in large cities and these have been used to articulate the "urban exodus" narrative. Facilitated by high air-travel connectivity, job density and spatial concentration of public-facing jobs, large cities became early global epicentres of COVID-19 infections during the early stages of the pandemic (Florida et al. 2021). Pre-coronavirus housing affordability and poor housing conditions were consistently identified as urban challenges in large cities. Coupled to these issues, lockdowns, social distancing, remote work and home schooling reportedly augmented the pressure for families living in small and crowded living spaces, to move out of cities in the look for more space and affordable housing (Hernández-Morales et al. 2020; Hughes 2020). Teleworking, increased familiarity and use of online shopping reduced the need for commuting and living in proximity to work and retail locations. Business closures removed the effervescence of urban entertainment, leisure and social spaces, and triggered a rapid spike in unemployment in many countries during 2020 as nonesential, public-facing work was suddenly paused (Falk 2020; Foley, Francis-Devine, & Powell 2020).

Enabled by automation and artificial intelligence, new digital technologies have greatly facilitated the transition to remote activities and arguably away from large cities during COVID-19 (Ting et al., 2020). Technologies, such as video conferencing, shared documents, instant messaging and cloud computing became instrumental in enabling remote work and education (Al-Maroof et al. 2020; Vargo et al., 2021). Virtual services, like video streaming and social media platforms offer access to some of the cultural effervescence and community that have consistently been an important factor drawing people to large cities (Glaeser, Kolko, & Saiz 2001; Harris & Todaro 1970). Online shopping platforms, such as Amazon and Ebay now provide an opportunity to buy and ship products from distant locations (Ting et al. 2020).

However, preliminary evidence suggests that population movements during the pandemic have been over relatively short distances. Evidence from the US and Spain suggests that most of the movement from large cities during the pandemic has been to their suburbs, as opposed to smaller, remote cities and towns (González-Leonardo et al., 2022; Hughes 2020). Yet, some city leavers also appear to have moved to neighbouring areas, second residences, holiday destinations and other cities (Kolko, Badger, & Bui 2021; Paybarah et al. 2020). In Australia, larger cities have been the primary destination for migration from other large cities, while the flow of people moving down the urban hierarchy has been much smaller (Davies 2021). In countries where anecdotal evidence exists, COVID-19 does not seem to have fundamentally altered the pre-existing national structure of the net internal balances of population movement. However, it seems to have accelerated relocation decisions that were already in motion pre-pandemic (Davies 2021; Kolko et al. 2021).

Persuasive cases have been made against headlines speculating about the end of cities. Past pandemics wreaked havoc and substantially influenced medical, cultural, political and urban design changes, but they have not dented the key role that cities play in society (Glaeser 2020; Reades & Crookston 2021). For instance, the Black Plagues of the 14th century killed one-third of the population in Europe and the Middle East (Pamuk 2007). The Cholera outbreaks of the 19th century decimated large cities across the world, including London, Paris, Moscow, Hamburg, New York and Madrid (Ali et al., 2015; Briggs 1961). Yet, large cities have continued to be important gravitational centres for population concentration.

Cities are critical engines of innovation, economic growth and prosperity. They enable the emergence of agglomeration economies. Concentration in cities facilitates the exchange of goods, knowledge, information and ideas by reducing transporta- tion and communication costs, offering abundant critical mass, and fostering strong firm linkages (Glaeser 2010). A fundamental ingredient underlying these benefits is the face-to-face interaction that can be fostered by urban agglomerations (Storper & Venables 2004). While routine, codified activities can be more easily communicated and performed virtually from remote locations, complex, innovative and less familiar tacit knowledge, tasks and ideas require face-to-face contact (Storper & Venables 2004). This need for face-to-face interaction comprises an essential reason why the proliferation of internet communication has not led to the spatial diffusion of urban agglomerations and "the death of distance", despite its capacity to enable complex ways of communication between distant locations (Fujita & Thisse 1996).

Additionally, rural and remote areas may lack the infrastructure and services needed to support incoming urban residents. These areas do not offer the vibrancy and sophistication of entertainment, cultural and convenient services that urbanities are used to. Telework is likely to remain a permanent way of interaction post-pandemic. Yet, poor broadband connectivity in rural and remote locations has remained a key challenge across most countries in the world (OECD 2020). Not all forms of work can be done remotely, including: high-touch, public facing work providing essential (e.g. healthcare and education services) and nonessential (e.g. restaurants, bars and clothing stores) services; essential, non-public facing work related to construction, infrastructure and maintenance; and knowledge-intensive activities requiring high level abstraction and cognitive capacity (i.e. teaching and networking) (Florida et al. 2021). Also online work fatigue has become a new phenomenon, reflected in the widespread use of term "Zoom fatigue" (Fosslien & Duffy 2020). Rather than full-time remote work, hybrid forms of work are thus more likely to outlast the pandemic, requiring flexibility to combine office and online presence. Such change may entail a need for reliable broadband connectivity and accessibility to employment centres.

Thus, while speculations during early stages of the pandemic pointed to an "urban exodus" as COVID-19 cases and deaths surged in large cities, emerging evidence suggests that the effects from the pandemic have reverberated through to the internal mobility system of countries across the world prompting residential relocations from large cities. Yet, such shocks are less likely to have led to a significant reconfiguration of the national mobility system. Rather, they may have accelerated existing mobility trends, with cities expected to bounce back and remain major centres of population attraction post-pandemic. Thus far, however, existing evidence remains largely anecdotal. We seek to offer some first evidence assessing the ways in which the British mobility system has weathered during the start of the pandemic to the reopening of the country's economy.

*2.2. Contemporary mobility patterns across the British urban hierarchy*

To determine the extent of change in mobility patterns during COVID-19, we review the pre-existing predominant trends of human mobility in the British system. Globally, the United Kingdom (UK) occupies an intermediate rank in terms of overall levels of mobility. Based on the 2011 UK Census, estimates indicate that 6.8 million individuals, or almost 11 in 100 people, changed their usual residential address in the last 12 months (Rowe et al., 2020). That is above the global average but well below countries, such as Iceland, Finland, the US and Australia (Bell et al. 2015). In the UK, while the share of short-distance migratory moves has been declining, most movement still occurs locally with approximately 30% of all residential changes taking place within 10km, while only less than 7% happen between distances of 50-200km and less 3% exceed 200km (Champion & Shuttleworth 2017).

A historical feature of the internal migration system in the UK for over the last half century has been counterurbanisation (Champion 1989). This process is characterised by population losses due to internal migration in major metropolitan areas and gains in smaller towns and rural areas. Over the last decade, a general decline in the size of net migration gains and loses across the national migration system has resulted in a weakening of the counter-urbanisation process (Lomax et al., 2014). This diminishing process has produced a pattern of spatial equilibrium in which migration inflows and outflows are closely balanced, resulting in minimal population redistribution across the national urban settlement (Rowe et al., 2019). Traditionally, acute net migration losses have been in London, the urban conurbation of the West Midlands, metropolitan districts in the North West, Glasgow, Edinburgh and Belfast (Lomax et al. 2014). Primary areas of net gain have been districts in the South West (especially Cornwall), and along the south coast and in the East of England (Lomax et al. 2014). Since 2008 following the global financial crisis, the number of metro-to-metro moves has hence replaced that of metro-to-nonmetro mobility as the predominant direction of migration flows in the UK (Lomax & Stillwell 2017). The number of nonmetro-to-metro moves has also increased exceeding the occurrence of nonmetro-to-nonmetro movement, although both of these types of flows have remained smaller than those occurring between metro areas, and from metro to nonmetro locales (Lomax & Stillwell 2017).

The importance of London in redistributing population represents a second relevant feature of the national internal migration network. Between 2010 and 2011, moves from and to London accounted for 15% of the total migration moves, or one in every seven moves, between local authority districts in the UK (Lomax & Stillwell 2017). London plays a key role as a social escalator region attracting young adults at rates which are higher than elsewhere in the country but recording significant losses of population due to migration across all other age groups as they step off the escalator and move away from London (Fielding 1992). London has thus consistently registered net migration losses, although these losses have lessened during the 2000s as a result of less acute outflows and greater inflows, particularly in inner boroughs (Champion 2015). Relevant in the context of the COVID-19 pandemic is the fact that net migration balances in London tend to correlate with economic cycles. Over the last four decades, London has registered the largest net migration losses during periods of national prosperity and lowest negative balances when economic conditions are less buoyant (Lomax & Stillwell 2017). While this evidence may suggest that moderate migration losses in London during the COVID-19 pandemic, as indicated earlier, reports have indicated large outflows from large dense cities. Next, we describe the data and methods used to assess the extent to which the existing patterns of population movement have been altered during the COVID-19 pandemic.

## 3 DATA

To capture population movements during the COVID-19 pandemic, we used anonymised aggregate mobile phone location data from Facebook users comprising 21 million observations for Great Britain and covering an 18-month period from March 23th 2020 to August 15th 2021. We used two data sets Facebook Movements and Facebook Population created by Meta and accessed through their Data for Good Initiative (https://dataforgood.facebook.com). The data sets are built from information from users who shared their location history. Before sharing the data, Meta applies three techniques to ensure privacy and anonymisation: random noise, spatial smoothing, and dropping small counts. First, a small undisclosed amount of random noise is added to ensure that precise location cannot be identified for small population counts in sparsely populated areas. Second, spatial smoothing is applied to produce a smooth population count surface using inverse distance-weighted averaging. Third, any remaining population counts of less than ten are removed from the final data set - see Maas et al. (2019) for details.

The Facebook Movements data set provides information on the number of Facebook users moving between and within locations. The Facebook Population data set provides information on the number of active Facebook users in a location at a given point in time. Both data sets provide daily population counts over three windows of eight hours: 00:00-08:00, 08:00-16:00 and 16:00-00:00 which are used to define the location of users. The location of users is defined as the place where they spent most of their time at a given time window (e.g. 00:00-08:00). Comparing the location of individuals between two temporal windows provides data on the number of people moving between locations. Both data sets include a baseline population count indicating the number of Facebook users moving between locations, or total number of Facebook users in a given location during a fixed baseline period. The baseline period is defined as an average of the population counts covering 45 days; that is, the 45 days prior to March 10th, 2020. The data sets also include a 'quality' score indicating the number of standard deviations by which the observed population count at a given time point differs from the baseline population count, highlighting statistically significant changes in population counts.

Facebook used the Bing Maps Tile System developed by Microsoft as a spatial reference framework to organise the data. It is a geospatial indexing system that partitions the world into tile cells in a hierarchical way, comprising 23 different levels of detail (Schwartz 2018). At the lowest level of detail (Level 1), the world is divided into four tiles with a coarse spatial resolution. At each successive level, the resolution increases by a factor of two. The Facebook mobility data we used are based on tiles at level of detail 12 which provides ground meters-to-pixel resolution of 38.2185 measured at the Equator. For Great Britain, that is a tile size of approximately 5.5-6km2.

We also used 1km2 gridded population data produced by Patias, Rowe, and Cavazzi (2019) derived from the 2011 UK Census to analyse population movement across the rural-urban continuum. We used gridded resident population counts to ensure consistency with the Facebook data. Based on an algorithm developed by Lloyd et al., (2017) (labelled PopChange), Patias et al. (2019) derived gridded population data from British censuses by calculating the correspondence between small area census geographies and 1km2 grids, and allocating population counts from each census area unit to its conforming 1km2 grids. We aggregated these data to Bing tile level 12 to match the Facebook data. This aggregation procedure was implemented via areal weighted interpolation using the package sf (Pebesma 2018) in the R environment.

Based on the resulting data, we used population density at the Bing tile level to capture the rural-urban continuum in the settlement hierarchy. This allows overcoming issues of comparability, spatial scale and measurement associated with the use of binary rural/urban classifications (Fielding 1989). We used deciles of population density to classify Bing tiles into ten discrete categories, combining tiles with similarly low population density and maximising the differentiation across moderate to high density population tiles. Figure 1 maps the resulting population density classes which tend to correspond to the Office for National Statistics (ONS) rural/urban classification (see Supplementary Material (SM) Figure 1). We preferred our population density classification as it provides a consistent definition of areas based on population density. The ONS rural/urban classifications for England, Wales and Scotland are generated independently based on different input data, and definitions of rural and urban - see SM Figure 2.

## 4 METHODS

The analysis involved two steps. First, we used area-based mobility metrics to measure the extent of change in mobility inflows, outflows and intraflows across the urban hierarchy at two key discrete points during the COVID-19 pandemic in the UK (i.e. after the first lockdown and after the implementation of the reopening). A detailed description of these events and the timeline of the COVID-19 pandemic in the UK is provided in Section 5.1. Second, we used statistical modelling to assess spatial and temporal variations in the intensity of mobility flows across specific origin-destination pairs of population density classes over the course of the pandemic.

We adopted an open and reproducible research approach based on the use of open software for mobility data analysis. We used data from accessible public and commercial sources following best practices in geographic data science (Brunsdon & Comber 2021). We produced an open data product (Arribas-Bel et al., 2021), including reproducible computational code to reproduce or extend our analysis, which is available for download as indicated in the Data Availability Statement.

[FIGURE 1 HERE]

4.1 *Area-based mobility metrics*

We measured changes in mobility inflows, outflows and intraflows across the urban hierarchy between two distinctive four-week periods during the course of the COVID-19 pandemic: (1) after the announcement of the first lockdown between March 23rd, 2020 and April 19th, 2020, and (2) after the implementation of the government's re-opening plan out of lockdown, or so so-called "freedom" day between July 19th, 2021 and August 16th, 2021. We measured changes in mobility during these periods, relative to pre-coronavirus levels during the baseline period as defined by Facebook; that is, a period covering 45 days before March 10th, 2020. Specifically, we computed the percentage change in mobility inflows, outflows and intraflows by population density class as follows:

$$I_{tc} = \left(\frac{\tilde{x}_{tc}}{\tilde{b}_c} - 1\right) * 100 \qquad (1)$$

where: $\tilde{x}$ corresponds to the median count for a specific type of population movement (i.e. inflows, outflows or intraflows; $b$ is the baseline median mobility count; $t$ relates to the two four-week periods after the first lockdown, or after the re-opening; $c$ refers to each population density classes, as defined in Section 3. A positive $I$ score indicates an increase in the extent of population movement relative to the baseline pre-pandemic period. A negative $I$ score represents a decrease in the extent of population movement relative to the baseline pre-pandemic period, while a zero score denotes no changes.

For the computation of Equation 1, we collapsed individual Bing tiles belonging to a same population density class and administrative area to create administrative-population-density geographical units. Intuitively, this spatial framework mitigates the influence of movement between adjacent tiles of a same population density class within the same administrative area, focusing on moves between different population density classes and administrative areas over relatively long distances. The resulting geographic boundaries cover 1,654 spatial units across Britain, with individual units averaging 60km2. The administrative units in the Facebook data correspond to the global administrative geographical classification system level 4 from the commercial company Precisely (www.precisely.com). These are the areas used by Meta to classify their mobility and population data.

4.2 *Modelling tile-to-tile mobility flows*

In a second stage, we used statistical modelling to understand differences in the intensity of mobility between tiles of different population density classes (as presented in Figure 1) and the extent to which the intensity of movement has evolved over time as the COVID-19 pandemic unfolds. In principle, the overall number of people moving between tiles of different population density classes could be estimated based on the original raw data. However, this approach could be misleading. Movement flows are not only influenced by the population nature of the origin and destination areas. Distance between locations, size, or socio-economic characteristics, among others, may also play a role as confounders, masking the true effect of population class. To unpick each of these and derive a "cleaner" estimate of the relevance of the nature of population, we opted for a regression modelling approach.

We adopted a spatial interaction framework. We used population flows between tiles by time window and day (as described in Section 3), and modelled them as a function of characteristics of the flow itself ($F_{ij}$), the origin ($Tile_i$) and destination ($Tile_j$) tiles, and the temporal nature of the flow ($T_w$). Crucially, we included indicator variables that capture the pair of population classes (Figure 1) of the origin and destination tile. In mathematical form:

$$\mu_{ijw} = \alpha + \underbrace{\sum_{IJ} \gamma_{IJ} + \beta_1 d_{ij} + \beta_2 q_{ijw}}_{F_{ij}} + \overbrace{\beta_3 Pop_i}^{Tile_i} + \underbrace{\beta_4 Pop_j}_{Tile_j} + \overbrace{D + H + Wk}^{T_w} \qquad (2)$$

where $\mu_{ijw}$ is the expectation of the flow of people from tile $i$ to tile $j$ in the time window $w$; $\alpha$ is an intercept; $\gamma_{IJ}$ is a series of indicator variables that reflect the pair of population density classes of a given origin $i$ ($I$) and destination tile $j$ ($J$), resulting in 99 pairs (10 classes x 10 classes minus one so it is not collinear with $\alpha$); $d_{ij}$ is the geographic distance between tiles $i$ and $j$; $q_{ijw}$ is a measure of the quality of the flow estimate provided by Meta-Facebook and related to the uncertainty behind the user count of the flow as described in Section 3; $Pop_{i,j}$ represents the population at the origin ($i$) and destination ($j$) tiles; $\beta_{1,2,3,4}$ are parameters to estimate in the model linking their respective covariates to $\mu_{ijw}$; $D$ is a trend tracking the day to which the flow relates to during the period in analysis; while $H$ and $Wk$ are indicator variables capturing day of the week (i.e. weekday or weekend) and hour window (i.e. 00:00-08:00, 08:00-16:00 and 16:00-00:00), respectively.

Our focus in Equation 2 is centred on $\gamma_{IJ}$. Controlling for all other variables in the model, these parameters capture the extent to which, the expected flow between a given origin-destination population density class pair of tiles (e.g. a high density origin to a low density destination) is systematically higher or lower than if it occurred between a baseline origin-destination population density class pair of tiles (e.g. a low density origin to a low density destination). Additionally, we standardised continuous variables ($d_{ij}$, $q_{ijw}$, $Pop_{i,j}$), $\alpha$ so that they can be interpreted as the expected flow on the first day ($D = 0$), during the first temporal window ($H = $ 00:00-08:00), on a weekday ($Wk = 0$), for the baseline origin-population population density class pair, when all the other variables are at their mean value. In this context, each $\gamma_{IJ}$ can also be seen as the "modulation factor" around that expectation associated with each pair of origin-destination classes. The baseline origin-destination population density class pair is the lowest population density class as origin and destination.

To model the flows, we used a count data regression model. Specifically, we fitted a generalised linear model (GLM) where the error term is assumed to be distributed following a Poisson distribution, with a flow expectation of ($\mu_{ijw}$) linked to the flow count ($F_{ijw}$) through a log link:

$$\log \mathbb{E}(F_{ijw}) = \mu_{ijw} \qquad (3)$$
$$F_{ijw} \sim Pois(e^{\mu_{ijw}}) \qquad (4)$$

The Poisson regression model (PRM) assumes that equidispersion; that is, the equality of mean and variance in the response variable (Cameron & Trivedi 2013). In practice, the equidispersion property is commonly violated because of overdispersion. This refers to the situation in which the variance exceeds the mean. When this occurs, the PRM may produce biased parameter estimates, causing the standard errors of the estimates to be underestimated, and compromising the statistical inference process regarding the extent, significance and direction of the influence of covariates (Hilbe 2011). To test for overdispersion, we used a regression-based test based on an auxiliary regression of the conditional variance generated from the predicted dependent variable on the conditional mean, without intercept term as described in Cameron and Trivedi (2005).

Following (Gelman & Hill 2006, pp.115-116), we used a quasi-PRM to address overdispersion in our response variable. This is one of the most common strategies to deal with overdispersion in count data models (Hilbe 2011, pp.158-159) Intuitively, this model adjusts the standard errors of the estimates to account for the extra dispersion in the data. To implement this, we estimated Equation 2 by using robust variance estimators. The number of active Facebook users were used as weight to account for the variability of observed count of population movement over time. This strategy is also used to mitigate for any potential biases regarding the variation in the observed number of active Facebook users changes over time across Britain.

We fitted Equation 2 using iteratively reweighted least squares (IWLS). We separately estimated models for individual months in our data, resulting in 18 sets of estimates. Our key aim is on estimates for α and $\gamma_{IJ}$, so that we focused on discussing the evolution of these estimates in a grid of line plots with ten rows and ten columns, each of them representing one of our population density classes (see Figure 1). The plot corresponding to the *I*-th row and *J*-th column displays the evolution of the parameter that tracks the intensity of population flows from tiles in population density class *I*-th to those in population density class *J*-th.

## 5 RESULTS

### 5.1 Timeline of COVID-19 in Britain

Understanding the timeline of government interventions during the COVID-19 pandemic in Britain is critical to the broader context within in which mobility patterns have occurred. A range of non-pharmaceutical interventions were implemented to reduce the spread of COVID-19 during the pandemic. Figure 2 shows the stringency index for the 18 months period covered by our data (March 2020 to August 2021). The stringency index is a composite measure developed by Hale, Webster, Petherick, Phillips, and Kira (2020), which captures the level of restrictions imposed by governments, ranging from 0 (loosest) to 100 (strictest).

[INSERT FIGURE 2 HERE]

Figure 2 identifies key five time intervals (Jennifer & Kirk-Wade 2021), with pink indicating increases in stringency and green denoting reductions in stringency. Darker colours points to greater changes in either of these directions. The first period corresponds to the first lockdown, which was announced on March 23th 2020 and imposed a "stay at home" order. During this period, only essential travel and outdoor physical activities were permitted, with the stringency index rising to 80. The second phase involves a period of partial reopening, with schools and non-essential shops being re-open on June 1 t 2020. Workers who could not work from home were also permitted to return to workplaces, but public transit was to be avoided. This partial re-opening was reflected in a moderately high stringency index ranging between 60 and 70. The third phase involves a period of fluctuating levels of stringency between October 14th 2020 until January 6th 2021 in response to changing COVID-19 cases and hospitalisation rates. It was marked by the introduction of a tier system with local lockdowns to locally contain centres of contagion, and then followed by a four-week second national lockdown between November 5th 2020 and December 2nd 2020, in response to rapidly spreading of COVID-19 across Britain. The fourth phase was characterised by a third national lockdown announced on January 6th 2021 to contain a rapid rise in COVID-19 cases due to the Delta variant. Levels of stringency during this phase peaked with a stringency index of 87. The fifth and final phase represented in Figure 2 displays the implementation of the government's plan out of the third national lockdown which sought to gradually lifted existing COVID-19 restrictions in four steps. Step 1 began on March 8th 2021 with the re-opening of schools and outdoor gatherings, followed by non-essential shops and outdoor venues (step 2), and subsequently indoor venues re-opening and large outdoor events (step 3). The re-opening strategy was completed on "freedom" day on July 19th when most of COVID-19-related legal restrictions were lifted.

### 5.2 Mobility patterns after the First Lockdown and "Freedom" Day

Next we analysed the percentage change in human mobility intensity at two distinctive points during the COVID-19 pandemic in Britain. As described in Section 4.1, Figure 3 shows percentage changes in inflows, outflows and intraflows between pre- pandemic mobility patterns during mobility and mobility patterns after the first lockdown, and mobility patterns post "freedom" day across population density classes. Positive percentage changes indicate increases in mobility intensity relative to the pre-pandemic period. Reductions in mobility are captured by negative scores and zeros denotes no change.

[INSERT FIGURE 3 HERE]

A key pattern from Figure 3 is an overall decline in mobility inflows and outflows across the urban hierarchy after the first national lockdown was enacted. On average, mobility declined by a 44% change. The sharpest declines are observed in highly densely populated areas, with declines exceeding 50%. Figure 3 also reveals a high level of variability in inflow and outflow mobility outcomes across the urban hierarchy. While high density population areas registered a consistent reduction in mobility intensity, less densely populated areas experienced more variable outcomes. These low-density areas recorded increases in inflows and outflows of up to 80%, compared to those mobility levels observed before the pandemic. These areas tend to comprise locales near and around national parks, such as the Pennines, Northumberland National Park, Peak District National Park, Galloway Forest Park and Nartmoor National Park, and coastal areas in South Wales. Instead, reductions in inflows and outflows tended to be larger and concentrate in highly density areas within large cities, including London, Birmingham and Manchester.

Declines in mobility inflows and outflows coincided with an overall rise in intraflows across the urban hierarchy, reflecting the effect of non-pharmaceutical interventions to contain the spread of COVID-19. Relative to pre-pandemic levels, intraflow movements increased by an average of 30%. Coupled with business and school closures and working from home, restrictions on outdoor activities and social gathering foster local mobility to access green spaces and essential services.

Following the full implementation of the government's exit strategy out of the third lockdown in July 2021, we observe increases in mobility intensity across the urban hierarchy. Mobility intensity bounced back closer to pre-pandemic levels, with less densely populated locations recording levels much closer to pre-pandemic patterns. Highly dense populous areas continued to record mobility levels between 15% to 20% lower than those occurring before the COVID-19 pandemic in 2020. Intraflows, however, reported a descending pattern across the urban hierarchy. Generally higher than pre-pandemic levels are observed in low density population areas, gradually declining as population density increases.

5.3 *Variations in mobility across the urban hierarchy*

Based on our modelling regression estimates, we then examine variations in the intensity of mobility across the urban hierarchy over the course of the COVID-19 pandemic. As described in Section 4.2, estimates were derived from a GLM Quasi-PRM using the number of moves as a function of a range of variables. In this section, we focus our analysis on estimates for our origin-destination class variables capturing variations in population density across origin and destination pairs. Estimates were derived from fitting individual regression models on pooled data sets for individual months. Full regression estimates and model diagnostics are reported in SM Figure 3 and Tables 1-2.

Figure 4 reports monthly regression estimates for the intercept (red line) and origin-destination class pairs (blue lines). The y-axis of Figure 4 represents origin population density classes, while destination classes are reported on the x-axis. Population density classes are encoded with numbers, with 1 indicating the least dense and 10 denoting the highest population density class. Individual plots are arranged according to these population density classes, with coefficients for the least dense population class at the top left of Figure 4, gradually increasing as we move towards the bottom and right. Each plot displays Poisson regression coefficients (y-axis) for individual origin-destination population density classes over time (x-axis). These coefficients capture changes in the intensity of mobility between origins and destinations of specific population density levels over the course of the COVID-19 pandemic. Coefficients for origin-destination pairs were derived from a fixed effects model, including a regression intercept as the baseline category. The first plot at the top-left corner of Figure 4 reports the regression intercept, indicating expected log mobility count between the least dense origins and destinations if all covariates are zero. This is interpreted as a baseline mobility count estimate, and the set of coefficients reported in the remaining plots are interpreted as deviations from this baseline estimate. Adding the regression intercept and a specific individual origin-destination pair produces an estimate of the overall effect relating to a particular class. We report individual coefficients for origin-destination pairs as they enable more easily determining the marginal effect of specific population density class on mobility flows. It enables isolating these marginal effects from overall effects which capture systematic nation-wide changes in human mobility patterns. Coefficients for origin-destination population class pairs indicate the expected log mobility count depending on their density class. A coefficient of 3 for origin class 10 and destination class 1, for instance, indicates that an additional 20 (= $e(3)$) people, on average, moved from the highest to the least population density class, relative to the baseline estimate. For visualisation purposes, we used a different y-axis scale for the regression intercept.

[INSERT FIGURE 4 HERE]

Recalling the national levels of mobility during the period in analysis is important to contextualise our estimates. Evidence so far points to reduced overall levels of mobility post announcement of the first national lockdown on March 23th 2020, bouncing back to pre-coronavirus levels following the implementation of the government's re-opening plan out of lockdown starting on March 8th 2021. Focusing first on the regression intercept, Figure 4 shows a positive coefficient for the least dense population origins and destinations fluctuating between 0 and 1 from March 2020 to February 2021, with a sharp drop in March 2021 and rise in April 2021, trending to 1 during the period from May to August. These patterns are consistent with low levels of mobility exchanges between the least dense population areas during periods of lockdown and partial re-opening during March 2020 to February 2021; a sudden reduction in mobility between these areas as the first stage of the re-opening plan out of lockdown was implemented in March 8th 2021; and, a marked increase as non-essential retail reopen in April 12th 2021.

Figure 4 reveals three key patterns. First, it reveals an overall predominant pattern of relatively high mobility involving low density areas (i.e. population density classes 1-5) as both origins and destinations during March 2020 to February 2021, rapidly declining during March to August 2021. This pattern suggests higher than average mobility levels from highly and less dense populated areas to low population density areas during periods of high stringency involving lockdowns, strict social distancing measures, school and business closures, as well as higher than average mobility levels from low density areas to high density areas. Declining patterns of mobility to and from low density areas post February 2021 coincide with the rolling out of the government's four-staged plan for re-opening from March 2021. The extent of this decline displays a clear spatial gradient of large reductions in inflows to low population density destinations, moving to more moderate declines in inflows to more densely populated locations. Changes in mobility relating to highly sparsely populated locales stand out displaying relatively high levels of movement, followed by a sudden decline when the third national lockdown was enacted.

A second key pattern relates to population exchanges between highly dense population areas (i.e. population density classes 6-10) across different population density classes. These areas tend to display relatively stable coefficients between March 2020 and August 2021. This suggests limited variations in the size of population movements in high density agglomerations during the pandemic, compared to population exchanges involving low density areas.

A third key feature involves population movements between population density areas of similar classes reported in the diagonal of Figure 4 . These exchanges tend to show a gradient of low mobility with limited variation in low density areas moving to higher levels of mobility in highly dense locations with a sudden rise in intensity following the implementation of the COVID-

19 exit strategy. This sudden increase in mobility intensity seems to have coincided with Stage 2 of the re-opening plan involving the resumption of nonessential retail activity and long-distance travel. These rises reflect increases in mobility exchanges between highly dense areas, but also higher intra-mobility intensity within these locations.

# 6 DISCUSSION AND CONCLUSION

## 6.1 *Key results and interpretation*

During the early stages of the COVID-19 pandemic, anecdotal reports of an "urban exodus" from big cities in various western societies emerged. Using smartphone application data from Facebook users, we sought to analyse the extent and durability of changes in human mobility patterns across the rural-urban hierarchy in Britain during the COVID-19 pandemic from March 2020 to August 2021. We found evidence of an overall and sustained decline in human mobility between areas during the enactment of non-pharmaceutical interventions between March 2020 and February 2021, bouncing back to pre-coronavirus levels following the roll out of the government lockdown exit strategy in March 2021. Declines in mobility between areas during high levels of stringency co-occurred with increases in mobility within areas, probably reflecting rises in active travel and e- forms of transport (Li et al. 2021). These patterns are largely consistent with evidence emerging from global data from Apple (Apple 2020) and Google mobility reports (Google 2020). We also showed that declines in mobility levels between areas in Britain varied markedly across the urban hierarchy, with the most densely populated areas experiencing the largest reductions; that is, a 60% decline from pre-pandemic levels. Declines in less populous areas were less acute and more variable, with some areas at intermediate levels of population density displaying rises in population of up to 80% in relation to pre-pandemic levels. These patterns are consistent with "the donut effect" used -by Ramani and Bloom (2021)- to describe population movement out of dense city centres to suburban areas around the largest twelve US cities during COVID-19.

We presented evidence of higher-than-average patterns of mobility from highly dense population areas to low densely populated areas as stringent non-pharmaceutical interventions were enacted and overall levels of mobility declined across Britain. This pattern is consistent with arguments of migration away from dense agglomerations as they became key early epicentres of COVID-19 infections and lost their urban vibrancy because of business, school closures, social distancing and lockdowns (Florida et al. 2021). This is in addition to pre-existing housing affordability and poor housing conditions, particularly in cities, like London (Edwards 2016). Collectively, these challenges seem to have exerted pressure to move out of dense cities as urban life was virtually shut down and small housing units had to be re-purposed into 24-hr multi-functional spaces to accommodate home schooling, telework and day-to-day activities (Capolongo et al. 2020; D'alessandro et al. 2020). However, we also presented evidence of higher-than-average mobility in the reverse direction (i.e. from low density areas to high density locations) and sustained high mobility between areas of relatively high population density. Taken together, our findings indicate that while patterns of population movement from densely populated agglomerations were higher than average, we found no evidence of COVID-19 leading to a "population exodus" from large cities.

We showed evidence of a systematic decline in mobility to low density areas, sustained mobility between high density areas, and a rise in mobility intensity between areas of similarly high population density levels as COVID-19 restrictions began to be lifted in Britain. These findings suggest that while COVID-19 generated shock waves leading to temporary changes in the pat- terns of population movement in Britain, the resulting vibrations have not significantly reshaped the prevalent existing structures in the national internal migration system. Large and dense urban areas are likely to remain key centre of population movement. Hybrid forms of working are likely to become widely adopted and predominant ways of interaction. Sparsely rural locations lack the infrastructure and services need to support hybrid working arrangements (OECD 2020). Poor broadband connectivity and deficient transport connectivity are likely to represent major challenges for these locations (OECD 2020). At the same time, urban areas already offer the required digital infrastructure. Economies agglomeration in dense urban spaces are likely to continue to facilitate and foster knowledge exchange, innovation and economic growth (Storper & Venables 2004). And in a similar way that the advent of internet communication has not led to the geographic dispersal of urban agglomerations (Rietveld & Vickerman 2004), it is difficult to envisage how the changes brought about by COVID-19 can trigger an urban exodus and redraw the national pattern of human population settlement.

*6.2 Limitations and future work*

Assessing the wider generalisability of our findings is challenging. Digital trace data are known suffer from biases and issues of representation reflecting differences in digital technology penetration, usage and accessibility (Rowe 2021). In the US, for instance, young adults particularly in the ages of 20 and 40 tend to be over-represented in Facebook data, while population over the age of 60 appear to be consistently under-represented (Ribeiro, Benevenuto, & Zagheni 2020). Yet, these biases do not seem to change our conclusions as these patterns of user representation concur with the mobility age schedule, with high intensity of mobility during young adult ages and gradually decreasing with ageing (Rogers, Raquillet, & Castro 1978). Thus, data are collected on the age groups that are more likely to move. Additionally, a recent UK-based study using the same mobility Facebook data set employed in our study demonstrated that they are strongly correlated with the spatial distribution of census and ONS mid-year population estimates (Gibbs et al. 2021). It also showed no systematic association between the percentage of Facebook users, and the average age, percent minority ethnic, population density, or index of multiple deprivation (Gibbs et al. 2021). Future work could extend our work triangulating other sources of data innovation, drawing a larger number of smartphone applications, as well as traditional data sources, like the 2021 British census when it becomes available. Such analysis could provide further ground-truthing of our results.

We analysed the spatial patterns of mobility across the rural-urban continuum. Future work is needed to establish the causes of the observed changes in the spatial direction of mobility during COVID-19. Understanding these causes can help anticipating long-term structural changes in mobility intensity extending beyond the pandemic. A combination of factors, including school shutdowns, business closures, social distancing, telework, employment density, housing space and affordability have been cited as key forces altering the pre-existing patterns of population movement during the pandemic, and triggering moves away from cities. While some of these factors have already dissipated as COVID-19 restrictions have been lifted, factors such as telework will most likely to endure the pandemic and become a main form of engaging with work (Florida et al. 2021). Assessing the extent to which companies can and will adopt remote work is key to understand the ways in which hybrid working can affect location decisions within and away from cities, in order to improve and design urban spaces by re-purposing office spaces and equip rural locations with needed digital infrastructure and transit connectivity.

**ACKNOWLEDGEMENTS**
We gratefully acknowledge funding from the Engineering and Physical Sciences Research Council (EPSRC) through the following projects: "ITINERANT: InequaliTies IN Experiencing uRbAn fuNcTion. Cuebiq data subscription" [EP/R511729/1]; and "InequaliTies IN Experiencing uRbAn fuNcTion (ITINERANT)" through The Alan Turing Institute [1162533].

**DATA AVAILABILITY STATEMENT**

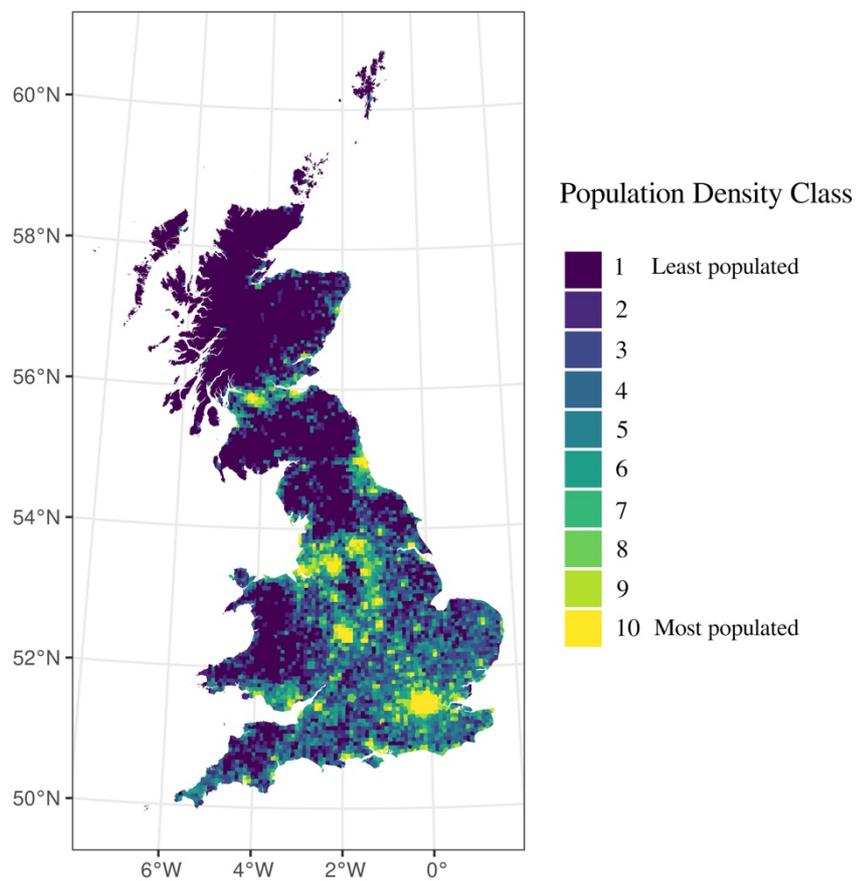

**FIGURE 1** Spatial distribution of Facebook tiles into population density classes. Class 1 includes the least densely populated, representing sparsely populated rural areas. Class 10 includes the most densely populated and highly urban agglomerations.

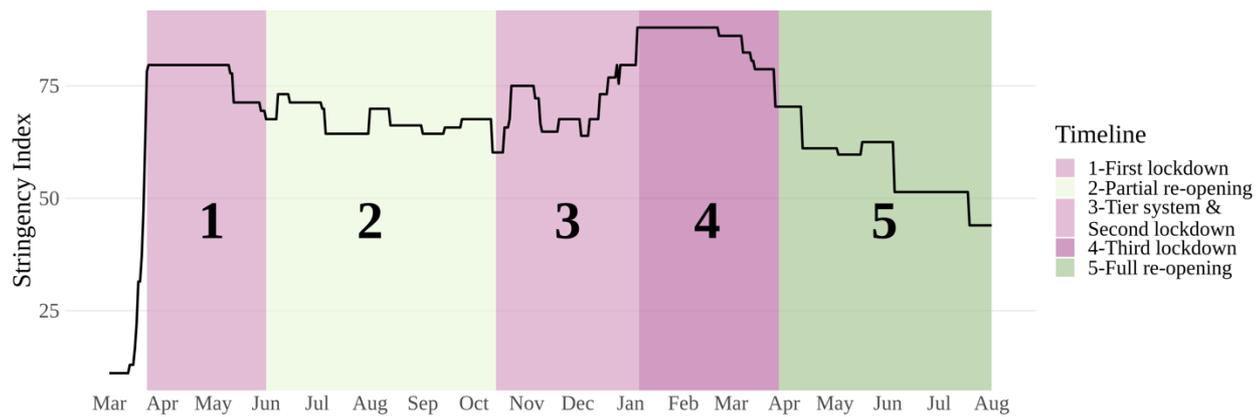

**FIGURE 2** Level of COVID-19 stringency measures in Britain, March 2020 to August 2021. The stringency index measures the level of non-pharmaceutical interventions to COVID-19, such as social distancing and lockdown measures and ranges from 0 to 100 (100 = strictest). The stringency index was sourced from COVID-19 government response tracker (https://www.bsg.ox.ac.uk/research/research-projects/covid-19-government-response-tracker) - see Hale et al. (2020) for more information.

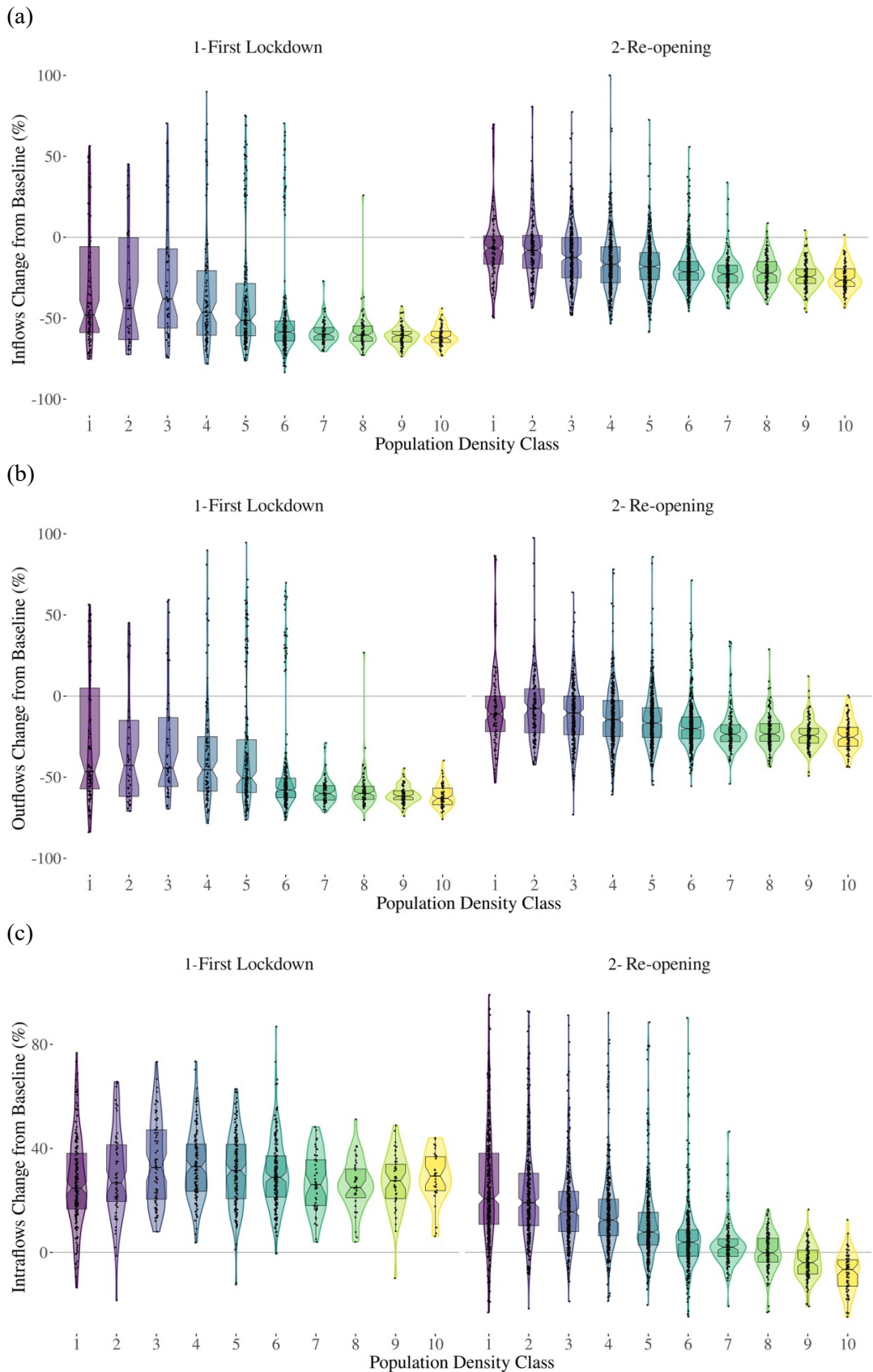

**FIGURE 3** Inflows (a), outflows (b) and intraflows (c) percentage change from pre-pandemic mobility standards after the first lockdown and "freedom" day across different population density classes. Positive values indicate % increase in flows compared to the baseline, negative values indicate % decrease, while zeros signal no change.

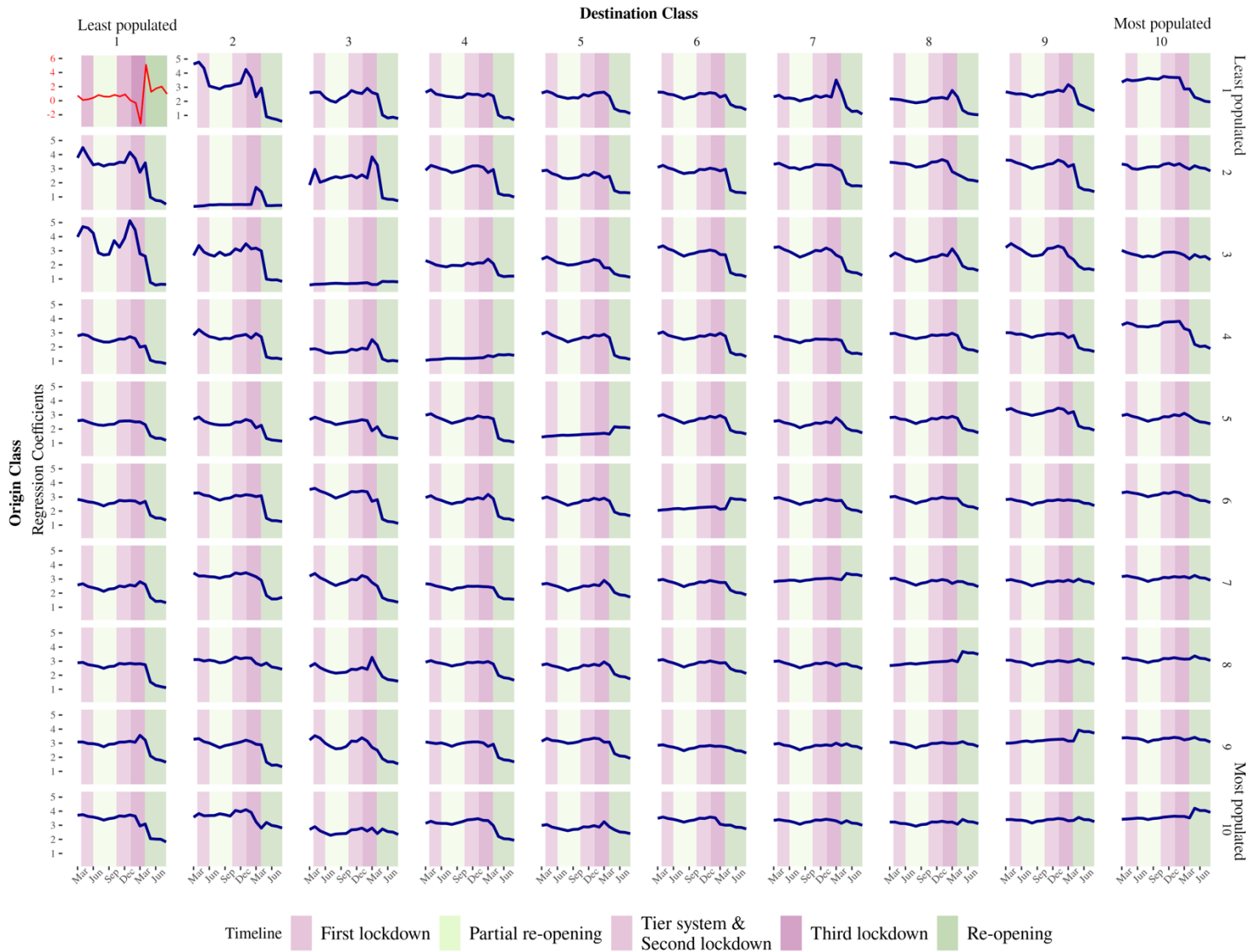

FIGURE 4 Poisson Regression Coefficients. The red line represents the regression intercept or reference category from separate monthly regression estimates. The regression intercept plot is reported on a different scale for the reasons discussed in the text. Blue lines represent the regression coefficients for each origin-destination population density pair, ranging from 1 (the least densely populated areas) to 10 (the most densely populated areas). Confidence Intervals (CIs) based at the 99% level of confidence are shown in grey. These intervals are however not visible as they are very small, with standard errors ranging from 0.0000007 to 0.005.

# Supplementary Material: Urban Exodus? Understanding Human Mobility in Britain During the COVID-19 Pandemic Using Facebook Data

Prepared for double-blind peer review

## Supplementary Tables

**Supplementary Table 1. Quasi-Poisson regression coefficients by month, March 2020-August 2021.** We report the regression model estimates included in Equation 2, excluding the estimates for origin-destination population density class pairs. Standard errors are provided in brackets.

|  | Distance | Origin Pop | Destination Pop | Day | Hour 08 | Hour 16 | Weekend | Quality Score |
| --- | --- | --- | --- | --- | --- | --- | --- | --- |
| 2020-03 | -3.15 (1e-05) | 0.179 (2e-06) | -0.037 (2e-06) | -0.146 (2e-06) | -3.859 (3e-06) | 0.145 (1e-06) | 0.055 (2e-06) | -0.121 (2e-06) |
| 2020-04 | 0.227 (3e-06) | 0.001 (0) | 0.16 (2e-06) | -0.054 (0) | 0.206 (1e-06) | 0 (0) | 0.253 (2e-06) | -0.04 (0) |
| 2020-05 | 0.183 (3e-06) | -0.189 (2e-06) | -0.11 (2e-06) | -3.527 (2e-06) | 0.144 (1e-06) | 0.066 (2e-06) | -0.125 (2e-06) | -3.304 (2e-06) |
| 2020-06 | -0.007 (1e-06) | 0.087 (2e-06) | -0.04 (0) | 0.22 (1e-06) | -0.003 (0) | 0.447 (2e-06) | -0.069 (0) | 0.228 (2e-06) |
| 2020-07 | -0.211 (4e-06) | -0.043 (2e-06) | -3.724 (2e-06) | 0.151 (1e-06) | 0.009 (2e-06) | -0.12 (2e-06) | -3.49 (2e-06) | 0.102 (2e-06) |
| 2020-08 | 0.067 (3e-06) | 0.064 (0) | 0.188 (1e-06) | -0.001 (0) | 0.361 (2e-06) | -0.063 (0) | 0.231 (2e-06) | -0.002 (0) |
| 2020-09 | -0.064 (3e-06) | -3.666 (3e-06) | 0.131 (1e-06) | -0.041 (2e-06) | -0.126 (2e-06) | -4.042 (3e-06) | 0.107 (2e-06) | -0.019 (2e-06) |
| 2020-10 | 0.152 (1e-06) | 0.234 (1e-06) | 0 (0) | 0.193 (2e-06) | -0.055 (0) | 0.193 (2e-06) | -0.001 (0) | 0.216 (2e-06) |
| 2020-11 | -3.877 (6e-06) | 0.168 (1e-06) | 0.042 (2e-06) | -0.102 (2e-06) | -4.197 (3e-06) | 0.176 (2e-06) | -0.035 (2e-06) | -0.114 (2e-06) |
| 2020-12 | 0.233 (2e-06) | 0 (0) | 0.218 (2e-06) | -0.034 (0) | 0.212 (1e-06) | 0.009 (0) | 0.2 (2e-06) | -0.027 (0) |
| 2021-01 | 0.191 (2e-06) | -0.152 (2e-06) | -0.189 (2e-06) | -3.711 (3e-06) | 0.151 (1e-06) | 0.092 (2e-06) | -0.092 (2e-06) | -3.266 (3e-06) |
| 2021-02 | 0 (0) | 0.092 (2e-06) | -0.085 (0) | 0.224 (1e-06) | -0.001 (0) | 0.447 (2e-06) | -0.01 (0) | 0.227 (2e-06) |
| 2021-03 | -0.225 (2e-06) | -0.065 (2e-06) | -3.515 (2e-06) | 0.154 (1e-06) | 0.011 (2e-06) | -0.145 (2e-06) | -3.305 (2e-06) | 0.096 (2e-06) |
| 2021-04 | 0.058 (2e-06) | 0.029 (0) | 0.21 (1e-06) | -0.001 (0) | 0.379 (2e-06) | -0.073 (0) | 0.227 (2e-06) | 0 (0) |
| 2021-05 | -0.048 (2e-06) | -3.727 (3e-06) | 0.146 (1e-06) | -0.066 (2e-06) | -0.118 (2e-06) | -3.787 (2e-06) | 0.104 (2e-06) | -0.027 (3e-06) |
| 2021-06 | 0.092 (1e-06) | 0.214 (1e-06) | 0.001 (0) | 0.24 (2e-06) | -0.049 (0) | 0.228 (2e-06) | -0.002 (0) | 0.23 (3e-06) |
| 2021-07 | -3.811 (4e-06) | 0.147 (1e-06) | -0.012 (2e-06) | -0.097 (2e-06) | -4.08 (3e-06) | 0.129 (2e-06) | 0.007 (2e-06) | -0.105 (3e-06) |
| 2021-08 | 0.236 (2e-06) | -0.003 (0) | 0.166 (2e-06) | -0.025 (0) | 0.212 (1e-06) | -0.013 (0) | 0.215 (2e-06) | -0.037 (1e-06) |

**Supplementary Table 2. Quasi-Poisson regression model diagnostics by month, March 2020-August 2021.** We report model diagnostics for our models, including measures of model fit (i.e. Akaike information criterion, Pseudo R Squared and Log-Likelihood), degrees of freedom and number of observations for each set of monthly estimates.

| Month-Year | Akaike | Pseudo R squared | Number of Obs. | Degrees of Freedom | Log-Likelihood |
|---|---|---|---|---|---|
| 2020-03 | 5.200832e+11 | 0.819 | 291439 | 291331 | -2.600416e+11 |
| 2020-04 | 1.447032e+12 | 0.826 | 831840 | 831732 | -7.235162e+11 |
| 2020-05 | 1.714408e+12 | 0.82 | 1046802 | 1046694 | -8.572042e+11 |
| 2020-06 | 1.860963e+12 | 0.809 | 1227192 | 1227084 | -9.304816e+11 |
| 2020-07 | 1.947047e+12 | 0.798 | 1405916 | 1405808 | -9.735233e+11 |
| 2020-08 | 1.822569e+12 | 0.798 | 1389682 | 1389574 | -9.112844e+11 |
| 2020-09 | 2.022856e+12 | 0.79 | 1462133 | 1462025 | -1.011428e+12 |
| 2020-10 | 2.084085e+12 | 0.79 | 1444114 | 1444006 | -1.042042e+12 |
| 2020-11 | 1.930502e+12 | 0.792 | 1320533 | 1320425 | -9.652510e+11 |
| 2020-12 | 1.825322e+12 | 0.796 | 1296650 | 1296542 | -9.126612e+11 |
| 2021-01 | 1.729239e+12 | 0.802 | 1173129 | 1173021 | -8.646193e+11 |
| 2021-02 | 1.596884e+12 | 0.801 | 1126290 | 1126182 | -7.984422e+11 |
| 2021-03 | 1.742603e+12 | 0.823 | 1290628 | 1290520 | -8.713015e+11 |
| 2021-04 | 1.730555e+12 | 0.808 | 1331641 | 1331533 | -8.652774e+11 |
| 2021-05 | 6.719025e+11 | 0.905 | 1287563 | 1287455 | -3.359513e+11 |
| 2021-06 | 7.480849e+11 | 0.9 | 1498527 | 1498419 | -3.740424e+11 |
| 2021-07 | 7.691850e+11 | 0.898 | 1523778 | 1523670 | -3.845925e+11 |
| 2021-08 | 3.720952e+11 | 0.891 | 720563 | 720455 | -1.860476e+11 |

# Supplementary Figures

**Supplementary Figure 1. Population density classes by decile and adjusted decile classes.** As described in Section 3, population density data were classified into ten discrete classes based on deciles. We generated an initial classification which was adjusted to reduce within-population density-class variability and maximise between-population-density-class variability. Specifically, areas belonging to the first four deciles were combined into a single class, the least densely populated category as very little differentiation exists across these classes as shown in the figure below. Areas belonging to the tenth decile were split into four classes based on the tenth decile's quartiles. The figure shows how the adjusted classes which, unlike official classifications (as discussed below), provide a consistent population density classification across the rural-urban continuum, and tend to better capture variations in the distribution of population densities in Britain. Figure 1 shows the spatial distribution of our final population density classification.

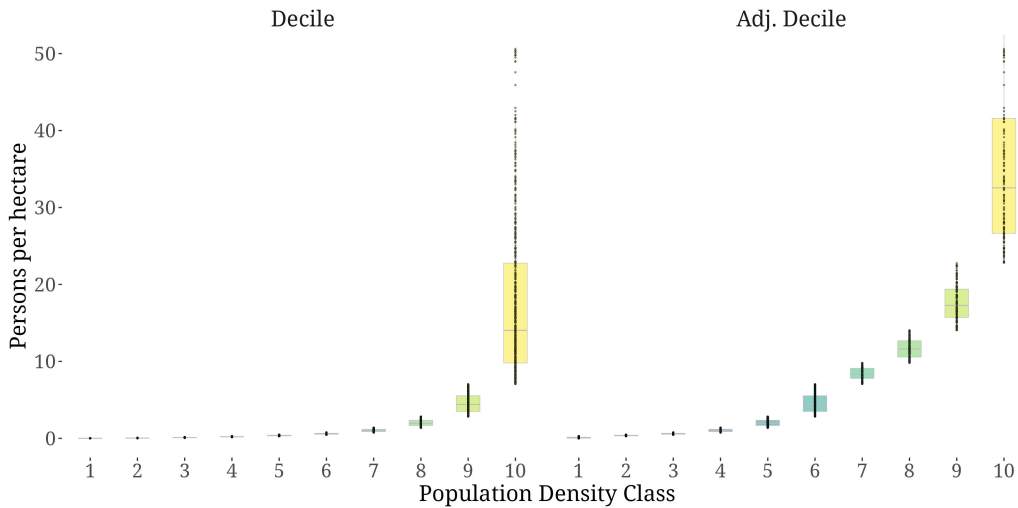

**Supplementary Figure 2. Rural-urban classification for Britain.** This figure displays the official rural-urban classification covering the British national territory for (a) England and Wales, and (b) for Scotland. These classifications were sourced from the Office for National Statistics and the Scottish Government, respectively. While these classifications offer a categorisation of the national territory across the rural-urban continuum, they are based on different definitions. Hence, we developed our own classification which maps to these categorisations offering a consistent framework.

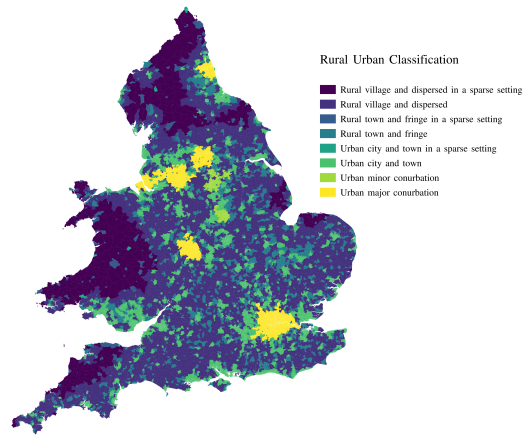

**(a)** Urban-Rural Classification in England and Wales.

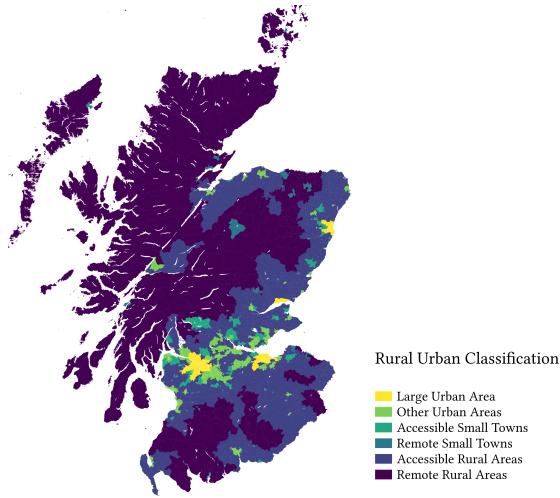

**(b)** Urban-Rural Classification in Scotland.

**Supplementary Figure 3. Quasi-Poisson regression model coefficients by month, March 2020-August 2021.** We plot the regression model estimates included in Equation 2, excluding the estimates for origin-destination population density class pairs. These estimates are reported in Figure 4. Detailed information on our model specification estimation can be found in Section 4.2. In the figure below, four colours are used to differentiate between model components: 1) distance; 2) population size at the origin and destination (i.e. OD population); 3) quality score, which is a metric provided by Facebook assessing the uncertainty in mobility flow estimates; and, 4) time, which refers to the temporal dynamics of our model involving a eight-hour window, day trend, and differences between weekdays and weekends.

The figure below shows a remarkably consistent association between time, OD population and quality score components and mobility flows over the course of COVID-19 pandemic, with coefficients ranging from -0.2 to 0.4. All coefficients show the expected direction of influence with larger populations at origin and destination, relating larger population movements. Coefficients for day, hours and weekend reflect small variations in the size of mobility flows over time, reflecting time trend linear changes, fluctuations in during the course of the day, and between weekdays and weekends.

As expected, distance shows a statistically significant negative coefficient, pointing to a negative association between the size of mobility flows and geographical distance. There is some variation in this association over time which tends to coincide with changes in stringency. Periods of high levels of stringency tend to correlate with greater coefficients for distance, reflecting the fact that people were deterred from moving over long distances and encourage to move within the local area.

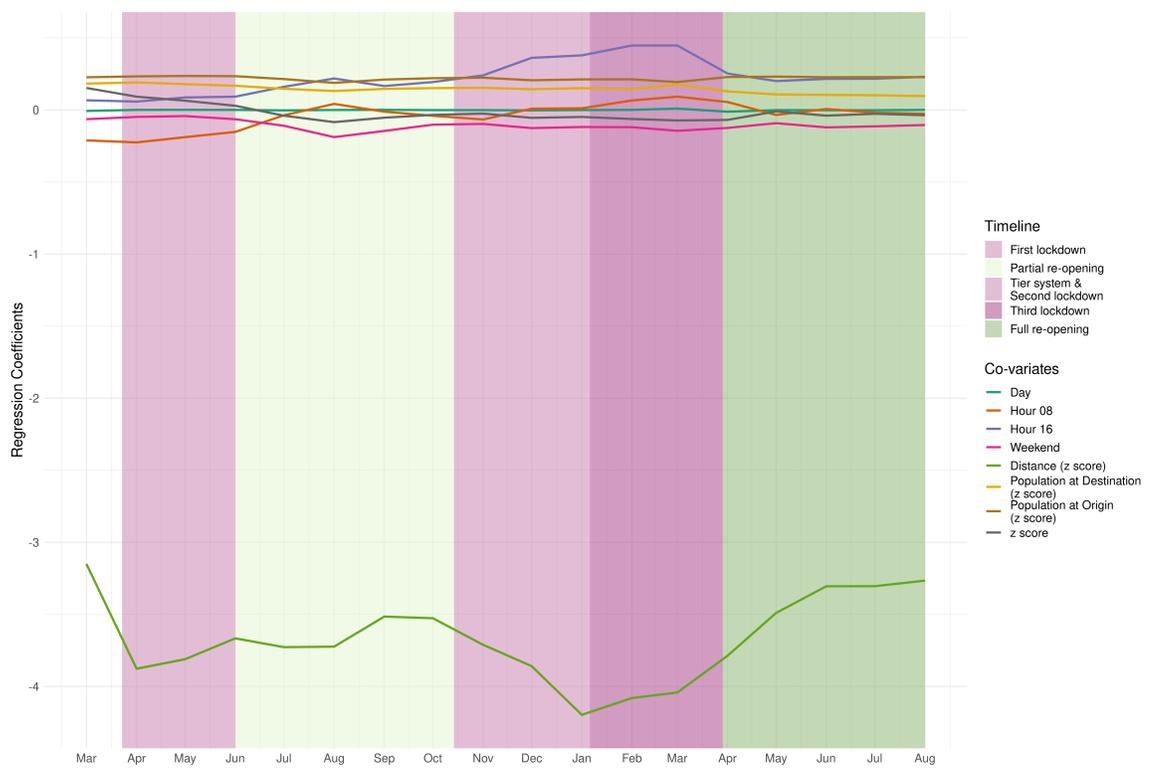